\documentclass{article}

\usepackage{arxiv}

\usepackage[utf8]{inputenc} % allow utf-8 input
\usepackage[T1]{fontenc}    % use 8-bit T1 fonts
\usepackage{hyperref}       % hyperlinks
\usepackage{url}            % simple URL typesetting
\usepackage{booktabs}       % professional-quality tables
\usepackage{amsfonts}       % blackboard math symbols
\usepackage{nicefrac}       % compact symbols for 1/2, etc.
\usepackage{microtype}      % microtypography
\usepackage{lipsum}		% Can be removed after putting your text content
\usepackage{multirow}
\usepackage{tabularx}
\usepackage{graphicx}
\usepackage{natbib}
\usepackage{doi}

\title{Self-supervised convolutional kernel-based handcrafted feature harmonization: Enhanced left ventricle hypertension disease phenotyping on echocardiography}

\author{ \href{https://orcid.org/0000-0003-1395-5474}{\includegraphics[scale=0.06]{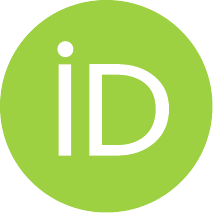}\hspace{1mm}Jina Lee} \\
	Department of Internal Medicine\\
	Graduate School of Medical Science\\
        Brain Korea 21 Project\\
        Yonsei University College of Medicine\\
        Seoul, Republic of Korea\\
	\texttt{qqwwdj@gmail.com} \\
	\And
	\href{https://orcid.org/0000-0003-2104-5905}{\includegraphics[scale=0.06]{orcid.pdf}\hspace{1mm}Youngtaek Hong} \\
	Ontack Health\\
	Seoul, Republic of Korea \\
	\texttt{hyt0205@gmail.com} \\
        \And
        \href{https://orcid.org/0000-0001-9791-9555}{\includegraphics[scale=0.06]{orcid.pdf}\hspace{1mm}Dawun Jeong} \\
	Department of Internal Medicine\\
	Graduate School of Medical Science\\
        Brain Korea 21 Project\\
        Yonsei University College of Medicine\\
        Seoul, Republic of Korea\\
	\texttt{jdwekdns1@gmail.com} \\
        \And
        {Yeonggul Jang} \\
	Ontack Health\\
	Seoul, Republic of Korea \\
	\texttt{jyg1722@gmail.com} \\
        \And
        {Jaeik Jeon} \\
        Ontack Health\\
	Seoul, Republic of Korea \\
	\texttt{jaeikjeon9919@gmail.com} \\
        \And
        {Sihyeon Jeong} \\
        Department of Internal Medicine\\
	Graduate School of Medical Science\\
        Brain Korea 21 Project\\
        Yonsei University College of Medicine\\
        Seoul, Republic of Korea\\
	\texttt{sihyeonjeong552@gmail.com} \\
        \And
        {Taekgeun Jung} \\
	Ontack Health\\
	Seoul, Republic of Korea \\
	\texttt{xorrms78@ontacthealth.com} \\
        \And
        {Yeonyee E. Yoon} \\
        Cardiovascular Center and Division of Cardiology \\
        Department of Internal Medicine \\
        Seoul National University Bundang Hospital \\
        Seongnam, Gyeonggi, Korea \\
	\texttt{yeonyeeyoon@gmail.com} \\
        \And
        {Inki Moon} \\
	Division of Cardiology\\
        Department of Internal Medicine\\
        Soonchunhyang University Bucheon Hospital\\
        Bucheon, Republic of Korea \\
	\texttt{rokstone4330@gmail.com} \\
        \And
        {Seung-Ah Lee} \\
	Ontack Health\\
	Seoul, Republic of Korea \\
	\texttt{seungah.lee@ontacthealth.com} \\
        \And
        {Hyuk-Jae Chang} \\
	  Division of Cardiology\\
        Severance Cardiovascular Hospital \\
        Yonsei University College of Medicine \\
        Yonsei University Health System \\
        Seoul, Korea
	\texttt{HJCHANG@yuhs.ac} \\
}

\renewcommand{\shorttitle}{LEE \textit{et al.}: Self-supervised convolutional kernel based handcrafted feature harmonization}

\hypersetup{
pdftitle={A template for the arxiv style},
pdfsubject={q-bio.NC, q-bio.QM},
pdfauthor={Jina Lee},
pdfkeywords={Self-supervised Learning, Harmonization, Echocardiography},
}

\begin{document}
\maketitle
\vspace{2cm} 
\begin{abstract}
Radiomics is a medical imaging technique that extracts quantitative handcrafted features from images to predict disease. The harmonization of these features ensures consistent feature extraction across various imaging devices and protocols. The methods for harmonization include standardized imaging protocols, statistical adjustments, and feature robustness evaluation. Myocardial diseases such as left ventricular hypertrophy (LVH) and hypertensive heart disease are diagnosed via echocardiography; however, variable imaging settings pose challenges. Thus, harmonization techniques are crucial for applying handcrafted features to disease diagnosis in such scenarios. Self-supervised learning (SSL) enhances data understanding within limited datasets and adapts to diverse data settings. ConvNeXt-V2 integrates convolutional layers into the SSL and exhibits superior performance in various tasks. This study focuses on convolutional filters within SSL for preprocessing images by converting them into feature maps for handcrafted feature harmonization. The method excelled in harmonization evaluation and exhibited superior LVH classification performance than existing methods.
\end{abstract}

% keywords can be removed

\section{Introduction}
Medical imaging has recently witnessed significant advancements, particularly in radiomics, which is a quantitative image analysis method that extracts numerous handcrafted features from radiological images \cite{gillies2016radiomics}. In radiomics, multiple handcrafted features are extracted from regions of interest (ROI), serving as inputs for prognosis prediction. Moreover, harmonization in handcrafted feature studies ensures that the extraction of quantitative imaging features is consistent and reproducible across imaging devices, protocols, and software. This is crucial because the same tumor imaged using different scanners or with different protocols may yield different radiomics features, which could lead to inconsistent or even contradictory results \cite{mali2021making}. Harmonization minimizes these discrepancies. Thus, several harmonization methods have been proposed to address these challenges. Note that the standardization of imaging protocols refers to the use of standardized imaging protocols that can reduce variability. For instance, specific guidelines for imaging parameters can be set for a particular study or multicenter trial. In this regard, ComBat harmonization \cite{johnson2007adjusting} is a statistical method initially developed for genomic data but has been adapted for universal harmonization techniques. ComBat adjusts for variations across multiple scanners and sites. For the model robustness, the handcrafted feature robustness to variations in imaging parameters and segmentation can be assessed. Only features stable across these variations can be used. Recently, convolutional neural network-based image synthesis methods have been proposed. For example, paired sample-to-sample synthesis DeepHarmony \cite{dewey2019deepharmony}, unpaired sample-to-sample synthesis CycleGAN \cite{zhu2017unpaired}, and Conditional VAE \cite{sohn2015learning} have been proposed. However, the harmonization of handcrafted features remains challenging.

Self-supervised learning (SSL) is a technique within unsupervised learning that enhances data understanding in a limited number of datasets. Among SSL techniques, ConvNeXt-V2 \cite{woo2023convnext} iis designed to incorporate convolutional layers into masked autoencoder \cite{he2022masked}. This approach have shown promising results in that the pre-trained ConvNeXt-V2 outperformed the supervised training in several downstream tasks, implying that the self-supervised convolution kernels can be generalized to apply to other downstream tasks. Thus, we focus on convolutional filters because of their generalized performance. In addition, we used a convolutional filter as a preprocessing filter to transform the image into a feature map. 

Myocardial disease occurs when damage to or death of heart muscle tissue occurs due to a lack of blood supply, often caused by a blocked coronary artery. Moreover, left ventricular hypertrophy (LVH) is a condition in which the muscle wall of the left pumping chamber of the heart (left ventricle) thickens (hypertrophy) \cite{stewart2018prognostic}. HHypertensive heart disease (HHD) and hypertrophic cardiomyopathy (HCM) represent LVH. In particular, chronic hypertension is the primary cause of HHD \cite{ommen20202020}. The heart must work harder to pump blood against increased resistance, leading to thickening of the heart muscle. In contrast, HCM is primarily a genetic condition caused by mutations in cardiac sarcomere proteins  \cite{unger20202020}. HCM can lead to LVH without high blood pressure or other cardiac stressors. HHD and HCM are diagnosed by echocardiography, and several studies have proposed distinguishing these diseases based on myocardial texture information. However, the challenge in echocardiography texture phenotyping from echocardiography images lies in the variability of imaging settings across different hospitals and even within the same facility \cite{yu2021artificial, kagiyama2020low}. This variability leads to substantial image differences, even for the same patient on the same day, owing to various parameters. Moreover, this variability poses a significant obstacle to implementing handcrafted features in disease diagnosis. Thus, harmonization techniques are necessary to mitigate such differences and build a clinically applicable machine learning model.

In this study, we aim to demonstrate the feature harmonization with the convolutional feature maps, and experimentally demonstrate the effectiveness of the proposed method with myocardial disease classification. This study is the first study to attempt to feature harmonization by utilizing the characteristics of convolutional neural networks.

The remainder of this paper is as follows. Section 2 described the methods to develop a harmonization filter that can perform well across different settings. Subsequently, Section 3 presents the results. The results are discussed in Section 4, including the study’s limitations and future work.

\section{Methods}
\label{sec:Methods}

\begin{figure}[htb!]
	\centering
	\includegraphics[width=150mm]{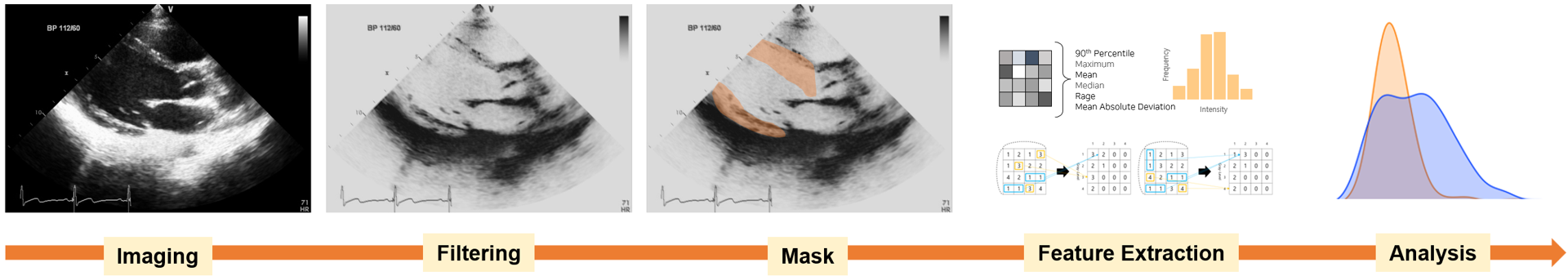}
	\caption{Harmonization Workflow.}
	\label{f1}
\end{figure}

We aimed to develop a harmonization filter that can perform well across different settings. In this regard, we implemented ConvNeXt-V2 to create a harmonized feature representation of echocardiography data. By training ConvNeXt-V2 on our echocardiography dataset with a masked autoencoder technique for self-supervised learning, we obtained a tuned model (ConvNeXtV2-Echo) that could learn highly representative features. We utilized the first convolutional layer of ConvNeXt-V2 as the 2D convolutional imaging filter, enabling us to generate filtered echocardiography images enriched with robust representative features. By applying the myocardium regional mask obtained from the original image to the filtered image, we could extract handcrafted features.

The workflow of our harmonization filter involves passing the original image through the filter, overlaying the myocardial mask obtained from the original image, and subsequently extracting and analyzing the handcrafted features from a specific region 
 (Fig. \ref{f1}).

\subsection{Data source}
\label{sec:Methods:sec1}

\begin{table}[ht]
\caption{Training Dataset}
\centering
\label{tab1}
\begin{tabularx}{\textwidth}{*{4}{>{\centering\arraybackslash}X}}
\hline
\multicolumn{4}{c}{\textbf{Manufacturer}} \\
GE      & Philips   & Siemens  & Toshiba  \\ \hline
1,706   & 359       & 78       & 5        \\ \hline
\end{tabularx}
\end{table}

\begin{table}[ht]
\caption{Downstream task Dataset}
\centering
\label{tab2}
\begin{tabularx}{\textwidth}{*{4}{>{\centering\arraybackslash}X}}
\hline
\multirow{2}{*}{\textbf{Disease}} & \multicolumn{2}{c}{\textbf{Manufacturer}}  & \multirow{2}{*}{\textbf{Total}} \\
& \multicolumn{1}{c}{GE} & \multicolumn{1}{c}{Philips} &                  \\ \hline
Normal                            & 356                    & 278                         & 634                             \\
HCM                               & 255                    & 46                          & 271                             \\
HHD                               & 161                    & 50                          & 211                             \\ \hline
\end{tabularx}
\end{table}

\subsubsection{Training dataset}
We utilized the Open AI Dataset Project (AI-Hub) (Ministry of Science and ICT, South Korean Government) \cite{AI-hub} to train ConvNeXt-V2. The dataset contains a compilation and refinement of echocardiogram data retrospectively acquired from multiple Korean institutions. This is a large open public dataset for adult transthoracic echocardiography. We employed this dataset for self-supervised training of ConvNeXtV2-Echo. For handcrafted feature extraction, we selected representative views: B-Mode Apical 2 Chamber view, Apical 4 Chamber view, Parasternal Long-Axis view, and Parasternal Short-Axis view with the deep learning method \cite{jeon2023echocardiographic}. The dataset incorporates four views from 2,148 individuals, amounting to 34,368 slides. Information on the imaging device and its manufacturer was obtained from the Dicom Header. In summary, 1,706, 359, 78, and 5 echocardiography images were acquired using the GE, Philips, Siemens, and Toshiba devices, respectively (Table \ref{tab1}).

\subsubsection{Downstream task dataset}
We retrospectively collected 1,116 B-Mode Parasternal Long-Axis view echocardiography DICOM datasets. This study was approved by the institutional review board (IRB:4-2020-1278). The datasets were acquired from four institutions: Sinchon Severance Hospital, Seoul National University Hospital, Bucheon Soonchunhyang University Hospital, and Hanyang University Hospital. We included 211 patients diagnosed with HHD and 271 patients with HCM. Data were acquired from various echocardiography devices manufactured by GE Healthcare and Philips Medical Systems (Table \ref{tab2}). The end-diastole (ED) and end-systole (ES) phases in the initial cardiac cycle were determined through manual selection guided by electrocardiogram signal analysis.

\begin{figure}[htb!]
	\centering
	\includegraphics[width=100mm]{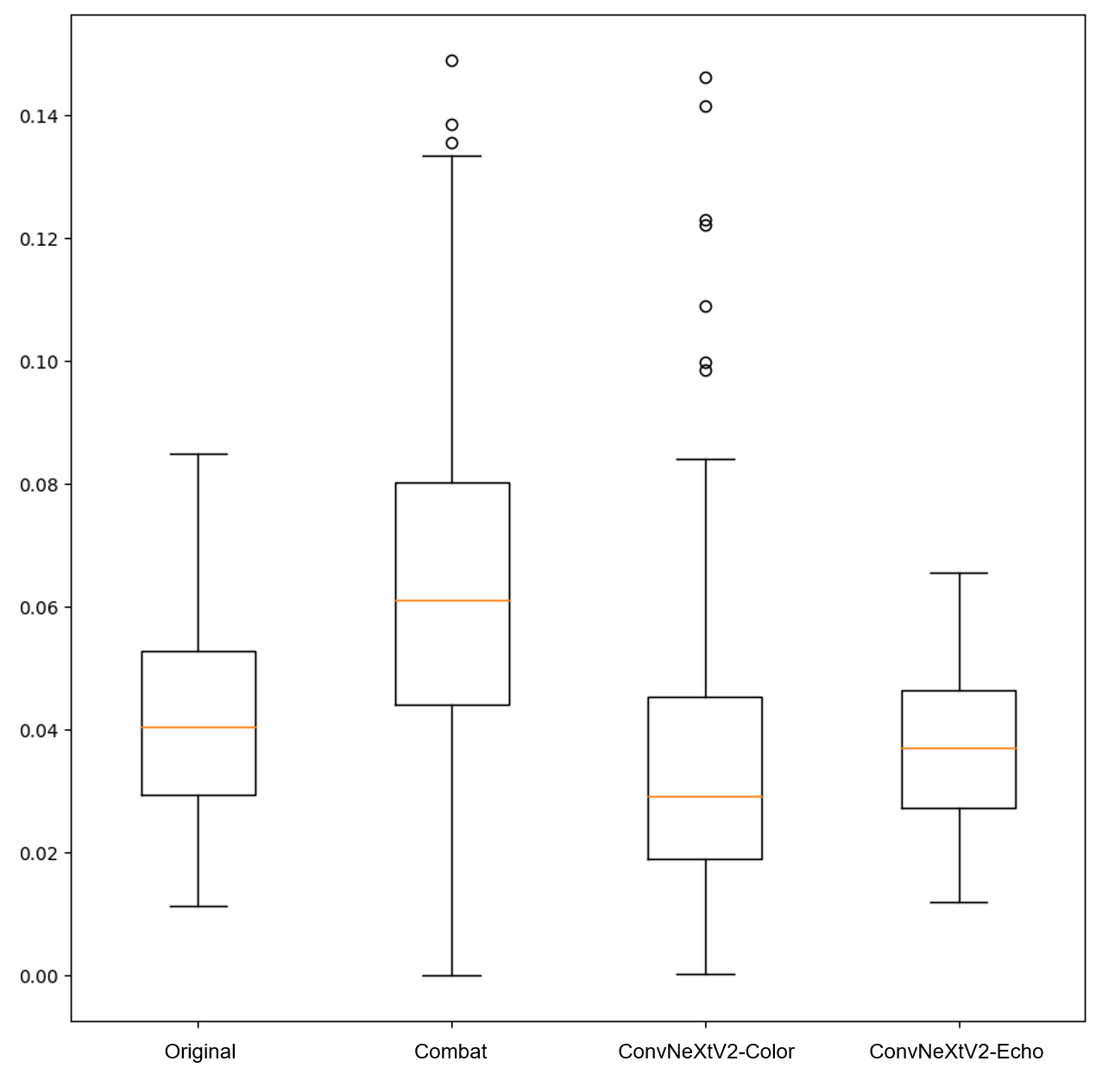}
	\caption{Boxplot of Jenson-Shannon divergence for original, Combat, ConvNeXtV2-Color, and ConvNeXtV2-Echo}
	\label{f2}
\end{figure}

\begin{figure}[htb!]
	\centering
	\includegraphics[width=150mm]{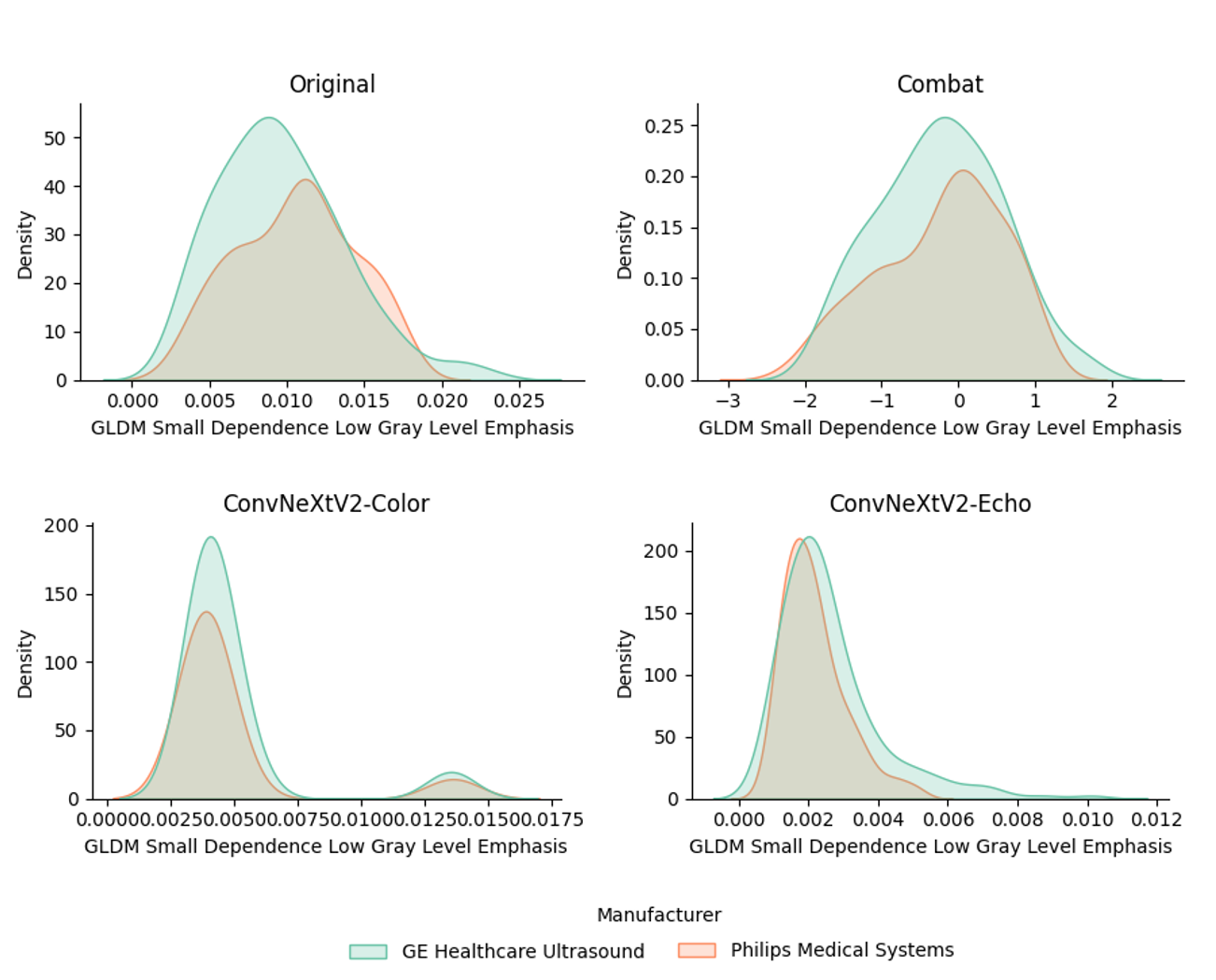}
	\caption{Histograms of example handcrafted feature (Small Dependence Low Gray Level Emphasis from GLDM class) of Original, Combat, ConvNeXtV2-Color, ConvNeXtV2-Echo}
	\label{f3}
\end{figure}

\begin{figure}[htb!]
	\centering
	\includegraphics[width=100mm]{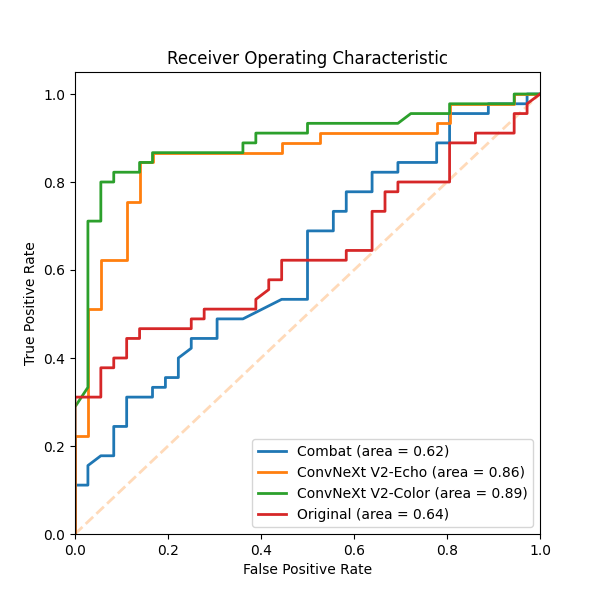}
	\caption{ROC curves for LVH binary classification}
	\label{f4}
\end{figure}

\begin{figure}[htb!]
	\centering
	\includegraphics[width=120mm]{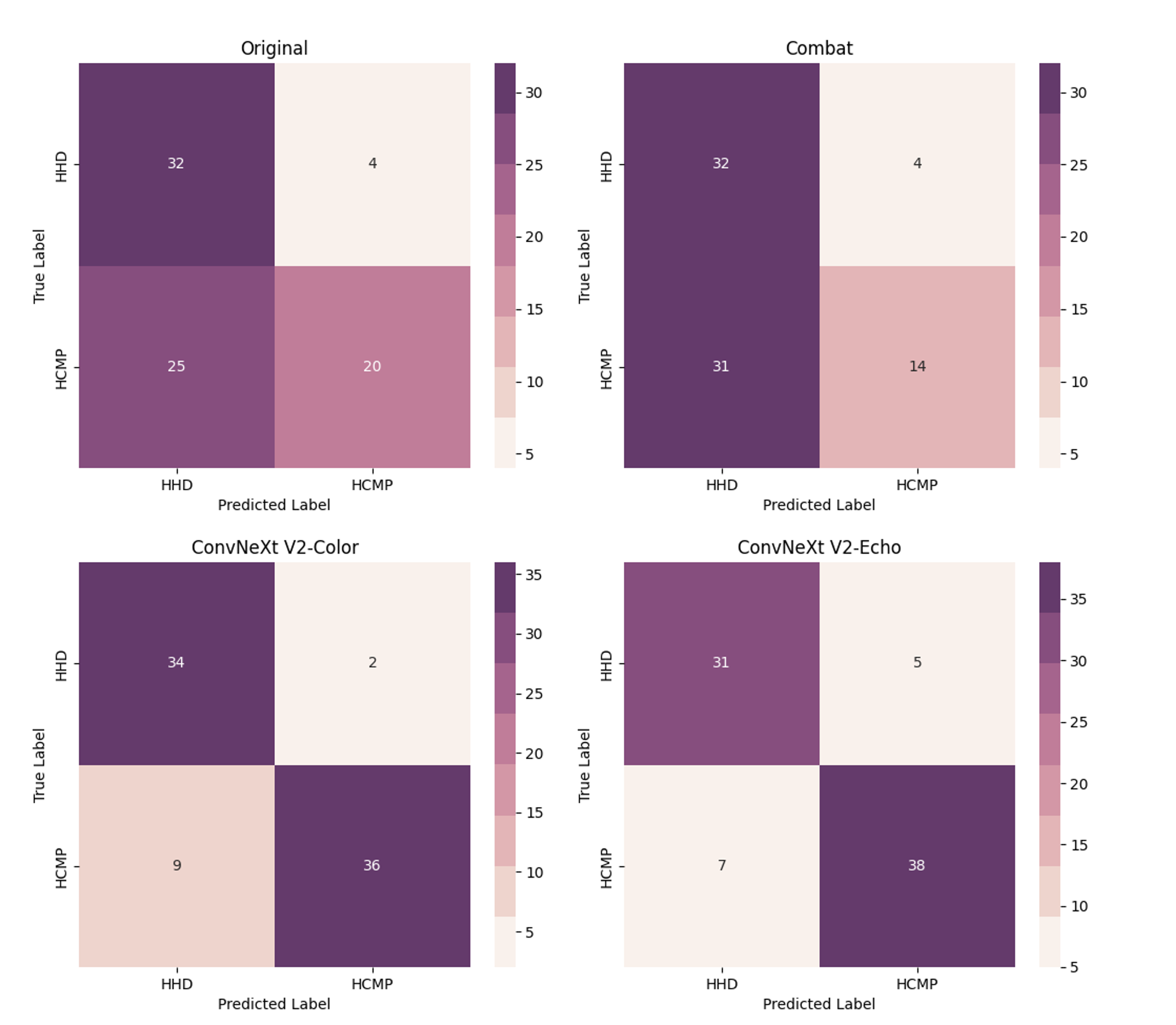}
	\caption{Confusion matrix for LVH binary classification}
	\label{f5}
\end{figure}

\subsection{Preparing harmonization filter}
\label{sec:Methods:sec2}

\subsubsection{Data preprocessing}
We applied two preprocessing steps to the dataset to prepare for the model training. First, given the inherent variability in echocardiography image sizes, we resized them to a uniform size of 448 × 448 pixels while preserving the original aspect ratio through padding. Second, we applied a mask to exclude regions outside the imaging region on the echocardiography. This process effectively eliminated nonessential information from the input to the model, ensuring that only relevant data were utilized.

\subsubsection{Neural Network Train}
We trained the base ConvNeXt-V2 model using our preprocessed echocardiography dataset. As the original ConvNeXt-V2 architecture was designed for 3-channel inputs for color images, we adapted it to accommodate grayscale echocardiography by configuring the input channel to 1.

In our trained model, we leverage the initial convolutional layer from the stem part of the model, which captures fundamental features, as the filter. This initial convolutional layer comprised 128 filters, each with dimensions of 4 × 4, and was operated with a stride of 4.

We transpose the learned weights and biases of the model into a convolutional layer with identical dimensions and filter counts to produce a filtered image maintaining the original dimensions. Employing a padding size of 1 and a stride of 1 for these layers ensured that the process would yield an output image of equivalent dimensions. Furthermore, we increase the input image size using an additional pixel. Finally, the final filtered image was derived by averaging the outcomes generated by all computed filters.

\subsection{Handcrafted features extraction}
\label{sec:Methods:sec3}

We extract handcrafted features from the myocardial area using a mask from a segmentation model. The mask in the myocardial region was acquired from cardiac ultrasound data using commercial software (Sonix Health, Ontact Health, Korea) \cite{kim2022fully}. Moreover, we used an open-source Python package, Pyradiomics 3.0.1 \cite{van2017computational}, to extract the features. The first step in feature extraction is the discretization of the grey values. We set the bin width to five to the features from the original intensity for discretization from the ROI. We extracted 93 radiomics, first-order, and higher-order statistical features. The details of the employed radiomic features are as follows: 18 first-order statistics features, 24 gray-level co-occurrence matrices (GLCM), 16 gray-level run-length matrices (GLRLM), 16 gray-level size zone matrices (GLSZM), 14 gray-level dependence matrices (GLDM), and 5 neighboring gray-tone difference matrices (NGTDM).

We extracted the features from the ED and ES phases of the cardiac cycle \cite{fukuta2008cardiac}. The utilization of the features extracted from the ED and ES varied depending on the evaluation method used. During Harmonization Analysis, the data extracted from each slide were treated as separate datasets. In contrast, the ED and ES data obtained from a single patient were considered as a single dataset for machine learning modeling.

\subsection{Harmonization analysis}
We employed the Jensen-Shannon divergence (JSD) as a metric to evaluate the performance of the harmonization techniques. The JSD is an information-theoretic method to measure the similarity between two probability distributions \cite{menendez1997jensen}. This metric is computed by averaging the Kullback-Leibler (KL) divergence values,  $\mathrm{KL}(P || Q)$ and $\mathrm{KL}(Q || P)$, between the probability distributions $P$ and $Q$. Essentially, the JSD helps quantify the difference in shape between two probability distributions, with smaller values indicating greater similarity in their shapes.

\subsection{Machine learning modeling}
\label{sec:Methods:sec4}
In our machine learning (ML) modeling, we randomly divided the LVH dataset, which comprised cases of HHD and HCM extracted from the downstream task dataset, into two sets, a training dataset and a test dataset, with an 8:2 ratio. After partitioning, we conducted feature selection exclusively on the training dataset. Subsequently, we trained a binary classification model using the selected features from the training dataset. The model performance was evaluated using a test dataset. In addition, we performed a Yeo-Johnson transformation \cite{yeo2000new} followed by z-score standardization to normalize the data. The Yeo-Johnson transformation effectively normalizes skewed data, aligning them in an optimal form resembling a normal distribution. The lambda parameter determines the degree of scaling, and the optimal lambda is obtained through maximum likelihood estimation. This transformation and subsequent normalization harmonized and standardized the data, mitigating the effect of outliers and skewed distributions. This process ensured that our analyses and predictions using the model were accurate and reliable. Our binary classification task involved discriminating between HCM and HHD. For this purpose, we used the XGBoost algorithm \cite{chen2016xgboost} as a regressor. To identify the most effective approach, we fine-tuned the search for optimal parameters of the technique.

\subsubsection{Feature selection}
\label{sec:Methods:sec5}
We employed the Boruta algorithm \cite{stoppiglia2003ranking, kursa2010feature} to select the important features. The Boruta algorithm addresses feature selection in the context of the random forest algorithm \cite{breiman2001random} and is essential to identify and select variables that significantly affect disease prediction accuracy. Within the Boruta algorithm, variables are systematically ranked based on their computed importance, facilitating the integration of the top-ranked features into classification modeling. The number of top-ranked features varied between the comparison groups. top-ranked features may vary among comparison groups.

\subsection{Statistical analysis}
\label{sec:Methods:sec6}
We utilized the area under the receiver operating characteristic curve (AUC) to evaluate the classification performance. Additionally, we calculated the Accuracy, Sensitivity, Specificity, and F1-score of each harmonization method for each disease. To establish the optimal cutoff values for each of the two diseases, we conducted preliminary calculations on the test set using the Youden index, J. Finally, we evaluated the predictive capabilities of the handcrafted features with harmonization using receiver operating characteristic (ROC) curve analysis. All analyses were performed using the Scikit-learn library, version 1.1.1.

\section{Results}
\label{sec:Results}

\subsection{Network training}
\label{sec:Results:sec1}
The experimental setup employed the AdamW optimization algorithm \cite{loshchilov2017decoupled} to train the ConvNeXt-V2 network. Specifically, we used a learning rate of 1×10-3 and set the parameters betas to 0.9 and 0.95. We adopted cosine annealing with a warm restart scheduler \cite{loshchilov2016sgdr} to control learning rate dynamics. The training regimen consisted of 800 epochs, and a batch size of 40 was used during the training process. Additionally, we incorporated a mask ratio of 0.2 and used a patch size of 32. The entire network was implemented in PyTorch version 1.13.1 and executed on two NVIDIA RTX A6000 GPUs.

\subsection{Harmonization result}
\label{sec:Results:sec2}

\begin{table}[ht]
\caption{Jensen-Shannon divergence Analysis with 15 bins of Histogram in Fig. \ref{f3}}
\centering
\label{tab3}
\begin{tabularx}{\textwidth}{*{3}{>{\centering\arraybackslash}X}}
\hline
Model            & Mean$\pm$STD   & Median \\ \hline
Original         & 0.0585$\pm$0.0746 & 0.448  \\
Combat           & 0.0735$\pm$0.0617 & 0.0643 \\
ConvNeXtV2-Color & 0.0638$\pm$0.0804 & \textbf{0.0339} \\
ConvNeXtV2-Echo  & \textbf{0.047$\pm$0.0367}  & 0.0399 \\ \hline
\end{tabularx}
\end{table}

We conducted a harmonization evaluation using a normal dataset extracted from a downstream task dataset. The normal dataset comprised 634 cases, of which 356 and 278 were acquired from GE and Philips, respectively. The harmonization assessed that handcrafted features from the two manufacturers did not exhibit statistically significant differences.

For the harmonization performance comparison, we employed the Combat Method and ConvNeXt-V2 pre-trained model. Combat \cite{johnson2007adjusting, fortin2018harmonization} is a method specifically devised to alleviate batch effects from diverse factors, such as acquisition equipment and environmental conditions. The purpose is to rectify these batch effects by leveraging the mean and variance within the designated batches. We implemented Combat on the normalized features to further reduce the variance attributed to using various manufacturers. In addition, we conducted comparisons with a pre-trained ConvNeXt-V2 model (ConvNeXtV2-Color) trained on a large-scale image dataset. Thus, the effectiveness of the proposed harmonization approach was assessed against a model trained on a dataset of diverse color images.

Furthermore, all data columns were discretized into 15 intervals, creating discrete probability distributions. Subsequently, we calculated the JSD values for all features to assess the variations between different manufacturers. ConvNeXtV2-Color and ConvNeXtV2-Echo achieved lower average JSD values than the original and Combat results (Table \ref{tab3}). Moreover, note that while ConvNeXtV2-Color filters yielded lower average JSD values than ConvNeXtV2-Echo filters, the latter displayed a more robust harmonization across all features without significant outliers (Fig. \ref{f2}). Contrastingly, the widely used Combat harmonization technique resulted in higher JSD values than the original data, suggesting that in the context of cardiac ultrasound data, this technique may worsen harmonization rather than improve it.

The results from the feature distribution in Fig. 3 also demonstrate that through convolutional filters, the features from both data groups were effectively harmonized to resemble a normal distribution shape.

\subsection{LVH classification and statistics}
\label{sec:Results:sec3}

\subsubsection{Feature selection}
The features were ranked based on their importance using the Boruta method. The results revealed that the original and Combat datasets had no features ranked 1, whereas ConvNeXtV2-Color had 16 features and ConvNeXt-Echo had 12 features ranked 1.

To conduct classification modeling, we exclusively utilized features with a rank of 1 from ConvNeXtV2-Color and ConvNeXt-Echo. We extracted features with ranks below 10 from the original and Combat results. Consequently, the original results contributed nine features, and the Combat results provided eight features for classification modeling.

\subsubsection{LVH classification}

\begin{table}[ht]
\caption{Performances of LVH classification model}
\centering
\label{tab4}
\begin{tabularx}{\textwidth}{>{\centering\arraybackslash}X *{5}{>{\centering\arraybackslash}X}}
\hline
Model            & Accuracy   & AUC   & Sensitivity   & Specificity   & F1-score  \\ \hline
Original         & 0.642      & 0.637 & 0.444         & \textbf{0.889}         & 0.580  \\
Combat           & 0.568      & 0.623 & 0.311         & \textbf{0.889}         & 0.444  \\
ConvNeXtV2-Color & \textbf{0.864}      & 0.\textbf{893} & 0.800         & 0.944         & \textbf{0.867}  \\
ConvNeXtV2-Echo  & 0.852      & 0.868 & \textbf{0.844}         & 0.861         & 0.864  \\ \hline
\end{tabularx}
\end{table}

The analysis revealed notable results in a binary classification task when distinguishing between patients with HCM and HHD. The modeling results of the handcrafted features harmonized with the ConvNeXt V2-Color filter achieved an AUC of 0.89, as shown in Fig. \ref{f4}. The ConvNeXt V2-Echo filter also showed a strong performance, with an AUC of 0.86. ConvNeXt V2-Echo achieved the highest sensitivity of 0.844. Moreover, the other performance metrics closely approximated those of ConvNeXt V2-Color, as outlined in Table \ref{tab4}. All convolutional filter-based methods outperformed the original dataset, whereas Combat demonstrated a lower performance than the original dataset. In Fig. \ref{f5}, the original dataset faces challenges in effectively distinguishing HCM. In contrast, the convolutional filter-based methods improved the prediction performance for HCM.

\subsection{Harmonization filter visualization}
\label{sec:Results:sec4}

\begin{figure}[htb!]
	\centering
	\includegraphics[width=150mm]{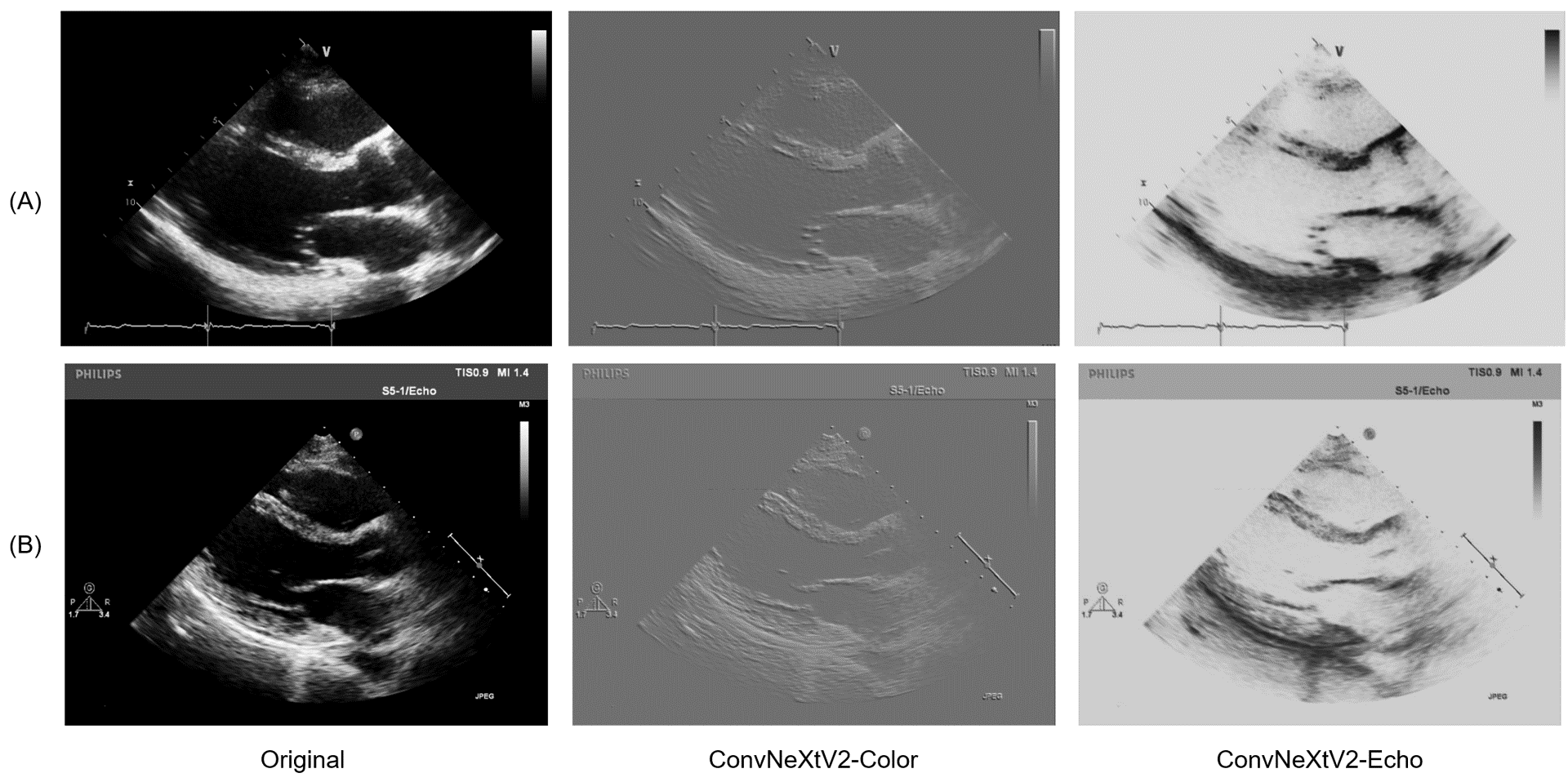}
	\caption{Visualization of result from Harmonization Filter (A) Echocardiography acquired from a device made by GE (B) Echocardiography acquired from a device made by Philips}
	\label{f6}
\end{figure}

\begin{figure}[htb!]
	\centering
	\includegraphics[width=150mm]{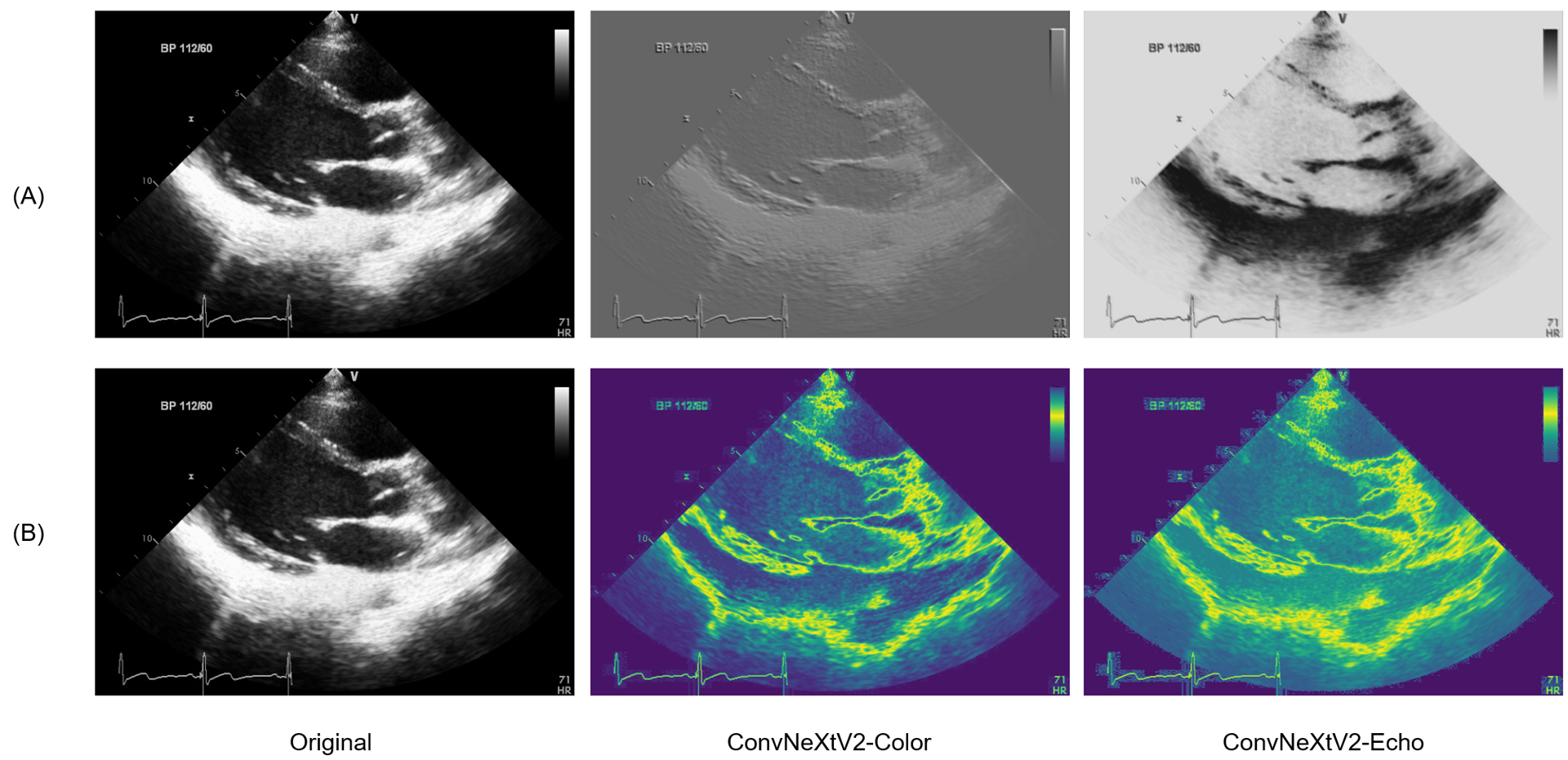}
	\caption{Visualization of result from Harmonization Filter (A) Original Image and Filtered Images (B) Difference with Original and Filtered Images}
	\label{f7}
\end{figure}

Fig. \ref{f6} shows the filtered images obtained using the convolutional filters ConvNeXtV2-Color and ConvNeXtV2-Echo. These filters were applied to the data collected from the GE and Philips sources following their respective convolutional filter styles.  In Fig. \ref{f7} suggests that ConvNeXtV2-Color exhibits pronounced distinctions, particularly along the boundaries of white regions.

\section{Discussion}
\label{sec:Discussion}

Analysis of radiological images with handcrafted features from radiomics has been crucial in numerous disease classification studies. However, obtaining handcrafted features from ultrasound images acquired across various devices, environments, and settings poses significant challenges to developing classification models. To overcome these limitations, we introduced a harmonization filter crafted through deep learning. To the best of our knowledge, this study is the first to use an SSL-based convolutional kernel as a harmonization filter.

Using SSL, we developed a filter that learned representative features from a given echocardiography dataset. The results of the harmonization process (Section \ref{sec:Results:sec2}) clearly showed that the individual features extracted from the filtered images were harmonized, and the overall feature distribution aligns much closer than the original feature distribution. 

Our study demonstrated a noteworthy enhancement in classification performance when applying the harmonized features obtained through the filter for the downstream task of LVH classification. In this context, we noted the following outcome: None were ranked 1 in the original set of features. In contrast, the Boruta algorithm ranked 1 a total of 12 handcrafted features from the filtered image. This observation indicates that various meaningful features were challenging to unearth owing to various noise factors. This serves as a clear reminder of the essential role of harmonization in the quest for meaningful feature exploration.

Moreover, the proposed harmonization technique sets a new standard for preprocessing handcrafted features extracted from echocardiography. Notably, the widely adopted harmonization method, Combat, showed a counterproductive effect on harmonization performance, as demonstrated by the JSD–Shannon divergence (JSD) and classification performance metrics. In contrast, our harmonization approach substantially enhanced the JSD values and classification performance.

The analysis showed that the ConvNeXtV2-Color, a model pre-trained on the ImageNet-1k database, slightly outperformed ConvNeXtV2-Echo. In particular, ConvNeXtV2-Color, trained on a vast dataset of over 1,000,000 color images, showed robustness and proficiency in extracting representative information from image data. Note that although our model was trained on a more modest dataset, its performance closely mirrored that of ConvNeXtV2-Color, indicating that ConvNeXtV2-Echo learned the representative information of the echocardiography dataset. Consequently, by acquiring a more extensive clinical dataset, ConvNeXtV2-Echo might increase its performance.

Despite the advancements in our study, several limitations were present. First, the sample size employed in the downstream task of the LVH classification requires an improved balance. The number of samples from the two manufacturers involved in the disease classification task differed by more than fourfold (GE 416, Philips 96). The substantial disparity in the number of samples from the two manufacturers restrained the assessment of the harmonization effectiveness. Second, external validation is essential to establish the generality of the harmonization techniques for clinical deployment. In future studies, we intend to address these limitations by acquiring additional datasets and further validating the efficacy of the proposed method.

\bibliographystyle{unsrtnat}
\bibliography{references}  %%% Uncomment this line and comment out the ``thebibliography'' section below to use the external .bib file (using bibtex) .

\begin{thebibliography}{27}
\providecommand{\natexlab}[1]{#1}
\providecommand{\url}[1]{\texttt{#1}}
\expandafter\ifx\csname urlstyle\endcsname\relax
  \providecommand{\doi}[1]{doi: #1}\else
  \providecommand{\doi}{doi: \begingroup \urlstyle{rm}\Url}\fi

\bibitem[Gillies et~al.(2016)Gillies, Kinahan, and Hricak]{gillies2016radiomics}
Robert~J Gillies, Paul~E Kinahan, and Hedvig Hricak.
\newblock Radiomics: images are more than pictures, they are data.
\newblock \emph{Radiology}, 278\penalty0 (2):\penalty0 563--577, 2016.

\bibitem[Mali et~al.(2021)Mali, Ibrahim, Woodruff, Andrearczyk, M{\"u}ller, Primakov, Salahuddin, Chatterjee, and Lambin]{mali2021making}
Shruti~Atul Mali, Abdalla Ibrahim, Henry~C Woodruff, Vincent Andrearczyk, Henning M{\"u}ller, Sergey Primakov, Zohaib Salahuddin, Avishek Chatterjee, and Philippe Lambin.
\newblock Making radiomics more reproducible across scanner and imaging protocol variations: a review of harmonization methods.
\newblock \emph{Journal of personalized medicine}, 11\penalty0 (9):\penalty0 842, 2021.

\bibitem[Johnson et~al.(2007)Johnson, Li, and Rabinovic]{johnson2007adjusting}
W~Evan Johnson, Cheng Li, and Ariel Rabinovic.
\newblock Adjusting batch effects in microarray expression data using empirical bayes methods.
\newblock \emph{Biostatistics}, 8\penalty0 (1):\penalty0 118--127, 2007.

\bibitem[Dewey et~al.(2019)Dewey, Zhao, Reinhold, Carass, Fitzgerald, Sotirchos, Saidha, Oh, Pham, Calabresi, et~al.]{dewey2019deepharmony}
Blake~E Dewey, Can Zhao, Jacob~C Reinhold, Aaron Carass, Kathryn~C Fitzgerald, Elias~S Sotirchos, Shiv Saidha, Jiwon Oh, Dzung~L Pham, Peter~A Calabresi, et~al.
\newblock Deepharmony: A deep learning approach to contrast harmonization across scanner changes.
\newblock \emph{Magnetic resonance imaging}, 64:\penalty0 160--170, 2019.

\bibitem[Zhu et~al.(2017)Zhu, Park, Isola, and Efros]{zhu2017unpaired}
Jun-Yan Zhu, Taesung Park, Phillip Isola, and Alexei~A Efros.
\newblock Unpaired image-to-image translation using cycle-consistent adversarial networks.
\newblock In \emph{Proceedings of the IEEE international conference on computer vision}, pages 2223--2232, 2017.

\bibitem[Sohn et~al.(2015)Sohn, Lee, and Yan]{sohn2015learning}
Kihyuk Sohn, Honglak Lee, and Xinchen Yan.
\newblock Learning structured output representation using deep conditional generative models.
\newblock \emph{Advances in neural information processing systems}, 28, 2015.

\bibitem[Woo et~al.(2023)Woo, Debnath, Hu, Chen, Liu, Kweon, and Xie]{woo2023convnext}
Sanghyun Woo, Shoubhik Debnath, Ronghang Hu, Xinlei Chen, Zhuang Liu, In~So Kweon, and Saining Xie.
\newblock Convnext v2: Co-designing and scaling convnets with masked autoencoders.
\newblock In \emph{Proceedings of the IEEE/CVF Conference on Computer Vision and Pattern Recognition}, pages 16133--16142, 2023.

\bibitem[He et~al.(2022)He, Chen, Xie, Li, Doll{\'a}r, and Girshick]{he2022masked}
Kaiming He, Xinlei Chen, Saining Xie, Yanghao Li, Piotr Doll{\'a}r, and Ross Girshick.
\newblock Masked autoencoders are scalable vision learners.
\newblock In \emph{Proceedings of the IEEE/CVF conference on computer vision and pattern recognition}, pages 16000--16009, 2022.

\bibitem[Stewart et~al.(2018)Stewart, Lavie, Shah, Englert, Gilliland, Qamruddin, Dinshaw, Cash, Ventura, and Milani]{stewart2018prognostic}
Merrill~H Stewart, Carl~J Lavie, Sangeeta Shah, Joseph Englert, Yvonne Gilliland, Salima Qamruddin, Homeyar Dinshaw, Michael Cash, Hector Ventura, and Richard Milani.
\newblock Prognostic implications of left ventricular hypertrophy.
\newblock \emph{Progress in cardiovascular diseases}, 61\penalty0 (5-6):\penalty0 446--455, 2018.

\bibitem[Ommen et~al.(2020)Ommen, Mital, Burke, Day, Deswal, Elliott, Evanovich, Hung, Joglar, Kantor, et~al.]{ommen20202020}
Steve~R Ommen, Seema Mital, Michael~A Burke, Sharlene~M Day, Anita Deswal, Perry Elliott, Lauren~L Evanovich, Judy Hung, Jos{\'e}~A Joglar, Paul Kantor, et~al.
\newblock 2020 aha/acc guideline for the diagnosis and treatment of patients with hypertrophic cardiomyopathy: executive summary: a report of the american college of cardiology/american heart association joint committee on clinical practice guidelines.
\newblock \emph{Journal of the American College of Cardiology}, 76\penalty0 (25):\penalty0 3022--3055, 2020.

\bibitem[Unger et~al.(2020)Unger, Borghi, Charchar, Khan, Poulter, Prabhakaran, Ramirez, Schlaich, Stergiou, Tomaszewski, et~al.]{unger20202020}
Thomas Unger, Claudio Borghi, Fadi Charchar, Nadia~A Khan, Neil~R Poulter, Dorairaj Prabhakaran, Agustin Ramirez, Markus Schlaich, George~S Stergiou, Maciej Tomaszewski, et~al.
\newblock 2020 international society of hypertension global hypertension practice guidelines.
\newblock \emph{Hypertension}, 75\penalty0 (6):\penalty0 1334--1357, 2020.

\bibitem[Yu et~al.(2021)Yu, Huang, Yu, Ma, Zhang, and Zhang]{yu2021artificial}
Fei Yu, Haibo Huang, Qihui Yu, Yuqing Ma, Qi~Zhang, and Bo~Zhang.
\newblock Artificial intelligence-based myocardial texture analysis in etiological differentiation of left ventricular hypertrophy.
\newblock \emph{Annals of Translational Medicine}, 9\penalty0 (2), 2021.

\bibitem[Kagiyama et~al.(2020)Kagiyama, Shrestha, Cho, Khalil, Singh, Challa, Casaclang-Verzosa, and Sengupta]{kagiyama2020low}
Nobuyuki Kagiyama, Sirish Shrestha, Jung~Sun Cho, Muhammad Khalil, Yashbir Singh, Abhiram Challa, Grace Casaclang-Verzosa, and Partho~P Sengupta.
\newblock A low-cost texture-based pipeline for predicting myocardial tissue remodeling and fibrosis using cardiac ultrasound.
\newblock \emph{EBioMedicine}, 54, 2020.

\bibitem[AI-()]{AI-hub}
URL \url{https://aihub.or.kr/aihubdata/data/view.do?currMenu=115&amp;topMenu=100&amp;aihubDataSe=realm&amp;dataSetSn=502}.

\bibitem[Jeon et~al.(2023)Jeon, Ha, Yoon, Kim, Jeong, Jeong, Jang, Hong, and Chang]{jeon2023echocardiographic}
Jaeik Jeon, Seongmin Ha, Yeonyee~E Yoon, Jiyeon Kim, Hyunseok Jeong, Dawun Jeong, Yeonggul Jang, Youngtaek Hong, and Hyuk-Jae Chang.
\newblock Echocardiographic view classification with integrated out-of-distribution detection for enhanced automatic echocardiographic analysis.
\newblock \emph{arXiv preprint arXiv:2308.16483}, 2023.

\bibitem[Kim et~al.(2022)Kim, Park, Jeon, Arsanjani, Heo, Lee, Moon, Yoo, and Chang]{kim2022fully}
Sekeun Kim, Hyung-Bok Park, Jaeik Jeon, Reza Arsanjani, Ran Heo, Sang-Eun Lee, Inki Moon, Sun~Kook Yoo, and Hyuk-Jae Chang.
\newblock Fully automated quantification of cardiac chamber and function assessment in 2-d echocardiography: clinical feasibility of deep learning-based algorithms.
\newblock \emph{The International Journal of Cardiovascular Imaging}, 38\penalty0 (5):\penalty0 1047--1059, 2022.

\bibitem[Van~Griethuysen et~al.(2017)Van~Griethuysen, Fedorov, Parmar, Hosny, Aucoin, Narayan, Beets-Tan, Fillion-Robin, Pieper, and Aerts]{van2017computational}
Joost~JM Van~Griethuysen, Andriy Fedorov, Chintan Parmar, Ahmed Hosny, Nicole Aucoin, Vivek Narayan, Regina~GH Beets-Tan, Jean-Christophe Fillion-Robin, Steve Pieper, and Hugo~JWL Aerts.
\newblock Computational radiomics system to decode the radiographic phenotype.
\newblock \emph{Cancer research}, 77\penalty0 (21):\penalty0 e104--e107, 2017.

\bibitem[Fukuta and Little(2008)]{fukuta2008cardiac}
Hidekatsu Fukuta and William~C Little.
\newblock The cardiac cycle and the physiologic basis of left ventricular contraction, ejection, relaxation, and filling.
\newblock \emph{Heart failure clinics}, 4\penalty0 (1):\penalty0 1--11, 2008.

\bibitem[Men{\'e}ndez et~al.(1997)Men{\'e}ndez, Pardo, Pardo, and Pardo]{menendez1997jensen}
ML~Men{\'e}ndez, JA~Pardo, L~Pardo, and MC~Pardo.
\newblock The jensen-shannon divergence.
\newblock \emph{Journal of the Franklin Institute}, 334\penalty0 (2):\penalty0 307--318, 1997.

\bibitem[Yeo and Johnson(2000)]{yeo2000new}
In-Kwon Yeo and Richard~A Johnson.
\newblock A new family of power transformations to improve normality or symmetry.
\newblock \emph{Biometrika}, 87\penalty0 (4):\penalty0 954--959, 2000.

\bibitem[Chen and Guestrin(2016)]{chen2016xgboost}
Tianqi Chen and Carlos Guestrin.
\newblock Xgboost: A scalable tree boosting system.
\newblock In \emph{Proceedings of the 22nd acm sigkdd international conference on knowledge discovery and data mining}, pages 785--794, 2016.

\bibitem[Stoppiglia et~al.(2003)Stoppiglia, Dreyfus, Dubois, and Oussar]{stoppiglia2003ranking}
Herv{\'e} Stoppiglia, G{\'e}rard Dreyfus, R{\'e}mi Dubois, and Yacine Oussar.
\newblock Ranking a random feature for variable and feature selection.
\newblock \emph{The Journal of Machine Learning Research}, 3:\penalty0 1399--1414, 2003.

\bibitem[Kursa and Rudnicki(2010)]{kursa2010feature}
Miron~B Kursa and Witold~R Rudnicki.
\newblock Feature selection with the boruta package.
\newblock \emph{Journal of statistical software}, 36:\penalty0 1--13, 2010.

\bibitem[Breiman(2001)]{breiman2001random}
Leo Breiman.
\newblock Random forests.
\newblock \emph{Machine learning}, 45:\penalty0 5--32, 2001.

\bibitem[Loshchilov and Hutter(2017)]{loshchilov2017decoupled}
Ilya Loshchilov and Frank Hutter.
\newblock Decoupled weight decay regularization.
\newblock \emph{arXiv preprint arXiv:1711.05101}, 2017.

\bibitem[Loshchilov and Hutter(2016)]{loshchilov2016sgdr}
Ilya Loshchilov and Frank Hutter.
\newblock Sgdr: Stochastic gradient descent with warm restarts.
\newblock \emph{arXiv preprint arXiv:1608.03983}, 2016.

\bibitem[Fortin et~al.(2018)Fortin, Cullen, Sheline, Taylor, Aselcioglu, Cook, Adams, Cooper, Fava, McGrath, et~al.]{fortin2018harmonization}
Jean-Philippe Fortin, Nicholas Cullen, Yvette~I Sheline, Warren~D Taylor, Irem Aselcioglu, Philip~A Cook, Phil Adams, Crystal Cooper, Maurizio Fava, Patrick~J McGrath, et~al.
\newblock Harmonization of cortical thickness measurements across scanners and sites.
\newblock \emph{Neuroimage}, 167:\penalty0 104--120, 2018.

\end{thebibliography}





% \end{thebibliography}

\end{document}